\documentclass[proceedings]{JHEP3}

\PrHEP{PrHEP hep2001}                   
\conference{International Europhysics Conference on HEP}                

\usepackage{epsfig}                   

\newcommand{\lsim}{\raisebox{-0.13cm}{~\shortstack{$<$ \\[-0.07cm] $\sim$}}~}
\newcommand{\gsim}{\raisebox{-0.13cm}{~\shortstack{$>$ \\[-0.07cm] $\sim$}}~}
\newcommand{\lsp}{\mbox{$\tilde\chi_1^0$}}
\newcommand{\om}{\mbox{$\Omega_{\tilde\chi_1^0} h^2$}}
\newcommand{\tb}{\tan\beta}
\newcommand{\ra}{\rightarrow}
\newcommand{\ee}{e^+e^-}

\title{Constraints on mSUGRA and  SUSY particle production at future $e^+e^-$ 
linear colliders}

\author{\speaker{Abdelhak Djouadi}$^1$                     
        Manuel Drees$^2$ and Jean-Loic Kneur$^1$\\         
        $^1$ LPMT, Universit\'e de Montpellier II, F--34095 Montpellier 
        Cedex 5, France. \\                                
        $^2$ Physik Department, TU M\"unchen, 
        D--85748 Garching, Germany.}                      

\abstract{We perform a complete analysis of the supersymmetric particle
spectrum in the Minimal Supergravity (mSUGRA) model. We show that present 
constraints on the Higgs boson and superparticle masses from collider 
searches and precision measurements still allow for large regions of the 
mSUGRA parameter space where some sparticles as well as the heavier Higgs 
particles, are light enough to be produced at the next generation of $e^+e^-$ 
linear colliders. An important part of this parameter space remains even 
when we require that the density of the lightest neutralinos left over from 
the Big Bang falls in the range favored by current determinations of the Dark 
Matter density in the Universe.}

\begin{document}

\section{Introduction}

Although other viable Supersymmetric (SUSY) of the Standard Model (SM) exist,
the Minimal Supergravity (mSUGRA) model has become the most frequently used
benchmark scenario for supersymmetry, and has been widely used to analyze the
expected SUSY particle spectrum and the properties of SUSY particles, and to
compare the predictions to available and/or expected data from collider
experiments.  Several global or partial analyses of the present theoretical and
experimental constraints on the mSUGRA model have been performed in the
literature.  In a recent paper [1], we have performed an independent
analysis of the SUSY particle is this model, taking into account theoretical
constraints and all available experimental information: searches for the MSSM
Higgs bosons and SUSY particles at the LEP and Tevatron colliders [2],
electroweak precision measurements [2], the radiative $b \to s\gamma$
decay, etc. Special attention was devoted to the implications of the
measurement of the anomalous magnetic moment of the muon recently performed at
Brookhaven [2], and to the $\sim 2\sigma$ evidence for a SM--like Higgs
boson with a mass $M_{\rm Higgs} \sim 115.6$ GeV seen by the LEP collaborations
[2]. We have also discussed the implication of requiring thermal relic
neutralinos to form the Dark Matter in the Universe [2]. We have then
discussed prospects for producing SUSY particles and the heavier Higgs bosons
of the MSSM at future high--energy $e^+e^-$ linear colliders with c.m. energies
around 800 GeV.  This talk summarizes the main results of this
analysis; for more details and for a complete set of references, we refer the
reader to the original work.  

\vspace*{-3mm} 

\section{The mSUGRA model and the calculation of the spectrum}

\vspace*{-3mm} 

We have performed our analysis in the constrained MSSM or mSUGRA model, where 
the MSSM soft breaking parameters obey a set of universal boundary conditions 
at the GUT scale, $M_{\rm GUT}\simeq 2.10^{16}$ GeV, so that the electroweak 
symmetry is broken radiatively. In this model, where the gauge couplings are 
unified at $M_{\rm GUT}$, one has only four continuous free parameters, and an 
unknown sign in addition to to the parameters of the SM:

\centerline{$
\tan \beta \ , \ m_{1/2} \ , \ m_0 \ , \ A_0 \ , \ \ {\rm sign}(\mu).$}

\noindent where $\tan \beta$ is the ratio of the vevs of the MSSM Higgs fields
$m_{1/2}, m_0$ and $A_0$, are respectively, the common soft--SUSY breaking
gaugino mass, scalar mass and trilinear couplings at the GUT scale, and $\mu$
is the higgsino mass parameter, the absolute value of which is determined by
the requirement of a proper electroweak symmetry breaking (EWSB). All the soft
SUSY breaking parameters at the weak scale are then obtained through
Renormalization Group Equations (RGE).

All results are based on the numerical FORTRAN code {\tt SuSpect} version 2.0 
[3], to which we refer for a more detailed description. The 
algorithm essentially includes:

-- RGE of parameters between the low energy scale and the GUT scale. For the
gauge and Yukawa couplings and gaugino masses, we use two--loop RGE. All the
one--loop SUSY threshold effects are implemented in the RG evolution via step
functions in the $\beta$ functions for each particle threshold.

-- Consistent implementation of EWSB.  Loop corrections (with all SUSY and
Higgs particles) to the effective potential are included using the tadpole
method. The SUSY parameters are frozen at the EWSB scale.  $\mu^2$ and $B\mu$
are determined from the minimization of the potential at this scale. Since
these parameters affect mass of some sparticles, this procedure has to be
iterated until stability is reached and a consistent value of $\mu$ is obtained.

-- Calculation of the physical (pole) masses of the Higgs bosons and the
sparticles including all the important ingredients. For instance, we include
the dominant radiative correction to the 3d generation fermion masses and to
all SUSY particles masses. The Higgs sector is treated in the effective
potential approach with RGE improved QCD corrections. In calculating the
masses, the procedure is iterated at least twice until stability is reached, in
order to take into account the (multi--scale) thresholds and the radiative
corrections.

In the numerical analyses we fix the MSSM parameters $\tan\beta$ given
at scale $M_Z$ as well as $A_0$ and the sign of $\mu$, and then
perform a systematic scan over the high energy mSUGRA inputs $m_0$ and
$m_{1/2}$. Given these boundary conditions, all the soft SUSY breaking
parameters and couplings are evolved down to the EWSB scale, which we choose 
to be the geometric mean of the two top squark masses, $M_{\rm EWSB} = 
(m_{\tilde{t}_1} m_{\tilde{t}_2})^{1/2}$. 

The program  allowed us to fairly reliably delineate the regions of the mSUGRA
parameter space which are still allowed by theoretical constraints [from a
proper EWSB breaking, neutralino LSP, non--tachyonic Higgs and SUSY particles,
etc..]

\section{Constraints on the mSUGRA parameter space}

\vspace*{-4mm} 

$\qquad $ {\it (i) Lower bounds on SUSY particle masses}:  
A wide range of searches for SUSY particles has been performed at LEP2
and at the Tevatron, resulting in limits on the masses of these
particles. The most important ones are due to the negative 
search of charginos, sleptons and third generation squarks at LEP2 and 
squarks and gluinos at the Tevatron. We therefore impose the following
bounds: $m_{\tilde\chi_1^+} \geq 104~{\rm GeV}, m_{\tilde{f}} \geq  100$ GeV 
with $\tilde{f}= \tilde{t}_1, \tilde{b}_1, \tilde{l}^\pm, \tilde{\nu}$ and 
$m_{\tilde{g}} \geq 300~{\rm GeV}, m_{\tilde{q}_{1,2}} \geq  260$ GeV with
$\tilde{q}= \tilde{u}, \tilde{d}, \tilde{s} , \tilde{d}$.

{\it (ii) Constraints from the Higgs boson masses}: In the SM, a 95\% CL lower 
bound has been set on the Higgs boson mass at LEP2, $M_{H^0} \geq 113.5$ GeV.
In the MSSM, this bound is valid in the decoupling regime where the pseudoscalar
$A$ boson is very heavy. For small values of $M_A$, a combined exclusion limit
of $M_h \sim M_A \geq 93.5$ GeV has been set. In the intermediate region an
interpolation has to be made. We have also studied the implications of the
$2.1 \sigma$ evidence for a SM--like Higgs boson with a mass $M_H = 115.6$ GeV 
seen by the LEP collaborations. In view of
the theoretical and experimental uncertainties, we interpreted this
result as favoring the range: $113 \, {\rm GeV} \, \leq \, M_h \, \leq 117 \, 
{\rm GeV}$.

{\it (iii) Constraints from electroweak precision observables:} Loops of Higgs
and SUSY particles can contribute to electroweak observables which have been
precisely measured at LEP, SLC and the Tevatron. The dominant contributions, in
particular $M_W$ and the effective angle $s_W^2$, enter via a deviation from
unity of the $\rho$ parameter which measures the breaking of the custodial
SU(2) symmetry.  In the MSSM, the dominant contributions are due to the 3d
generation $(\tilde{t}, \tilde{b})$ and $(\tilde{\tau}, \tilde{\nu})$ weak
iso--doublets, which we have required these contributions to stay below the
acceptable ($2\sigma$) level of $ \Delta \rho (\tilde{f}) \leq 2.2 \cdot
10^{-3}$.

{\it (iv) The $b \to s \gamma$ constraint:} Another observable where SUSY
particle contributions might be large is the radiative flavor changing decay $b
\to s\gamma$, the branching ratio of which has been measured to be ${\rm BR}(b
\to s \gamma) = (3.37 \pm 0.37 \pm 0.34 \pm 0.24^{+0.35}_{-0.16} \pm 0.38)
\cdot 10^{-4}$, including theoretical errors. In our analysis, we will use the
most up--to--date determination in the MSSM of the $b \to s \gamma$ decay rate
including NLO QCD corrections and allow the branching ratio to vary in the
$2\sigma$ range: $2.0 \times 10^{-4} \leq {\rm BR}(b \to s \gamma) \leq 5.0
\times 10^{-4}$.

{\it (v) The contribution to the muon $g-2$:} Recently, the Muon $(g-2)$
Collaboration has reported a new measurement of the anomalous moment of the
muon: $(g_\mu-2) \equiv a_\mu^{\rm exp} = 11\, 659\, 202\, (14)(6)
\, 10^{-10}$, which differs from the predicted SM average value by 2.6$\sigma$.
We interpret the discrepancy as being a SUSY contribution (chargino--sneutrino
and neutralino--smuon loops) of $11 \cdot 10^{-10}
\leq a_\mu^{\rm SUSY} \, \leq 75 \cdot 10^{-10}$.

{\it (vi) Cosmological constraints}: We have analyzed the contribution of the
$\chi_1^0$ particles, which are the lightest SUSY particles, to the
(normalized) overall matter density of the Universe $\Omega_{\tilde\chi_1^0}
h^2$ ($h\sim 0.5$ is the Hubble constant).  The $\chi_1^0$, is neutral, weakly
interacting, massive and absolutely stable since R-parity is conserved in
mSUGRA, and is therefore a good candidate for the cold Dark Matter. Recent
evidence suggests that $\Omega_{\tilde\chi_1^0} h^2 \simeq 0.2$ and we define
$0.1 (0.025) \leq \Omega_{\tilde\chi_1^0} h^2 \leq 0.3 (0.5)$ as the
(conservative) cosmologically favored region.  The calculation of the relic
density is made using standard assumptions and includes all the annihilation
channels, with a proper treatment of the $s$--channel poles, as well as
co--annihilation with gauginos, sleptons and top squarks.  

The main outcome of this part our analysis can be summarized as follows: 

-- There are large areas of the $(m_{1/2}, m_0)$ parameter space which
are still allowed by present experimental constraints. In particular,
for large enough values of $\tb$, the bound on the lightest $h$ boson
mass, $M_h \gsim 113$ GeV, does not place too severe constraints. If
$\mu>0$, which is favored by the $(g_\mu-2)$ anomaly, the constraint
from the radiative decay $b \to s\gamma$ are not severe 
even for large values of $\tb$. In fact, if $A_0 = 0$ it is always
superseded by the Higgs boson mass constraint, but for $A_0 = -1$ TeV
the $b \ra s \gamma$ constraint can be more severe. 
Precision electroweak measurements are easily accommodated. 

-- For $\tb \gsim 10$ and small values of the coupling $A_0$, the requirement
of 113 GeV $\lsim M_h \lsim 117$ GeV favors moderate values of the gaugino
mass, $m_{1/2} \lsim 500$ GeV, leading to relatively light chargino and
neutralino states, $m_{\tilde\chi_1^\pm} \sim m_{\tilde\chi_2^0} \sim 2
m_{\tilde\chi_1^0} \lsim 400$ GeV. For large (and negative) values of $A_0$,
which lead to a strong mixing in the stop sector, $M_h$ in this range can be
accommodated in large regions of the parameter space even for rather small $\tb
(\sim 5)$ values. In this case $\tilde t_1$  can be rather light, if the
parameters $m_0$ and $m_{1/2}$ are not too high. The range of $m_{1/2}$ favored
by the LEP Higgs evidence strongly depends on the exact value of $M_t$,
calling for a more precise determination of this parameter.

-- The $(g_\mu-2)$ excess, which can be accommodated in the MSSM only
if $\mu >0$, typically gives a stronger upper bound on $m_0$ than the
requirement $M_H = 115 \pm 2$ GeV. For $\tb\sim 40$, $m_0$ and
$m_{1/2}$ values below $\sim 600$ GeV [and slightly above $\sim 300$
GeV] are needed, implying again relatively light electroweak gaugino
and slepton states. However, the value of this upper bound increases
roughly proportional to $\tb$, so that at $\tb = 60$, $m_0$ as large
as 1.0 (1.6) TeV can be accommodated at the 1 (2) $\sigma$ level. 

-- For small and moderate $\tb (\lsim 40)$ the requirement that the
density of the lightest neutralinos accounts for the Dark Matter density in 
the Universe is
very constraining indeed. In this case most of the region where $\om$
is ``naturally'' in the interesting range is excluded by the Higgs
mass constraints, which requires SUSY breaking masses above those
preferred by Dark Matter calculations. Only a small band in the region
with a relatively light bino--like neutralino and relatively light
$\tilde{\tau}$ survives. In addition, there are ``exceptional''
regions: a narrow strip in the $\tilde \tau_1 \lsp$ co--annihilation
region near the boundary where the $\tilde{\tau}_1$ slepton is the
LSP, and a strip in the focus point region at large $m_0$ and small
$m_{1/2}$ values where neutralinos and charginos are relatively light
and have large higgsino components. Requiring in addition $M_h = 115
\pm 2$ GeV and a SUSY interpretation for the $(g_\mu - 2)$ anomaly
removes most of these ``exceptional'' regions with acceptable relic
density.  On the other hand, for large values of $\tb (\gsim 50)$, the
area of the $(m_0, m_{1/2})$ parameter space favored by cosmology
extends significantly due to the opening of the pseudoscalar
$A$--boson pole. This allows to fit all the requirements [$M_h$,
$(g_\mu-2)$ and the DM constraint] in a somewhat larger area of the
$(m_0, m_{1/2})$ parameter space. 

-- In spite of the strong constraints on mSUGRA obtained by taking seriously
all the positive indications for SUSY it is still not possible to give tight
limits on any one single parameter.  We found overlap regions with $5 \leq \tb
\leq 60$, 0.1 TeV $\lsim m_0 \lsim$ 1.5 TeV, 160 GeV $\lsim m_{1/2} \lsim$ 550
GeV.  Furthermore, allowing for a large negative $A_0$ plays an important role
in extending the allowed region to smaller values of $\tb$. In fact, the
allowed $(m_{1/2}, m_0)$ region plane could be further extended by considering
more $A_0$ choices. 

\vspace*{-2mm}

\begin{figure}[htbp]
\hspace*{1.5cm}{\large $m_0$}\\[-1.5cm]
\begin{center}
\hspace*{-.2cm} \epsfig{figure=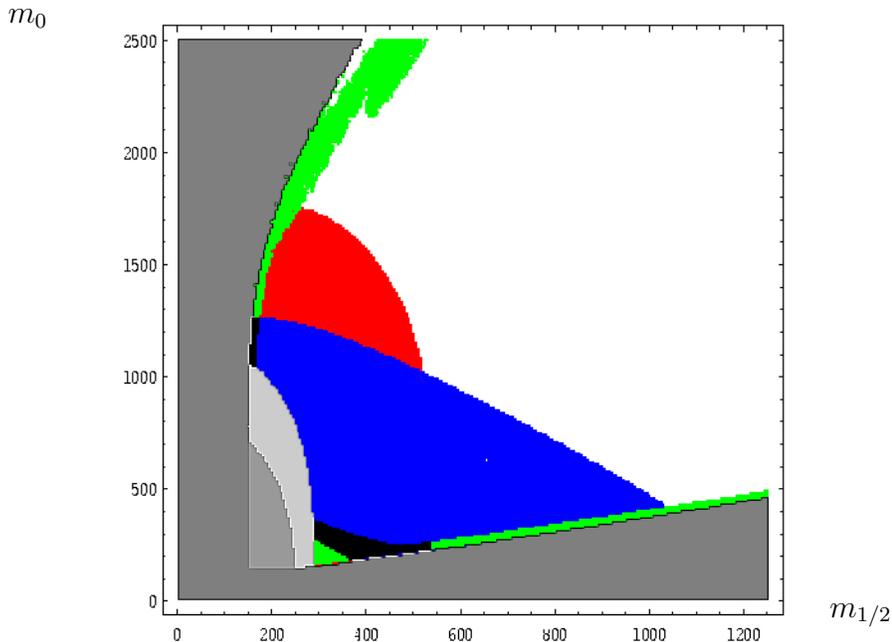,width=8.5cm,angle=-90}\\
\end{center}
\vspace*{-1cm}
\hspace*{12.3cm} {\large $m_{1/2}$}
\vspace*{1mm}
\caption[]{Constraints on the $(m_{1/2}, m_0)$ mSUGRA plane for 
$\tan \beta=40$, $A_0=0$ and sign$(\mu)>0$. The grey areas 
are those excluded by the requirement of EWSB and limits on SUSY 
particle masses (darker grey), BR($b \to s \gamma)$ (medium grey) and 
$M_h > 113$ GeV (light and dark grey). The colors are for the ``evidence" for
the $h$ boson (red), the $(g_\mu-2)$ (blue) and Dark Matter (green).}
\vspace*{-.2cm}
\end{figure}
\begin{figure}[htbp]
\hspace*{1.5cm}{\large $m_0$}\\[-1.5cm]
\begin{center}
\hspace*{-.2cm} \epsfig{figure=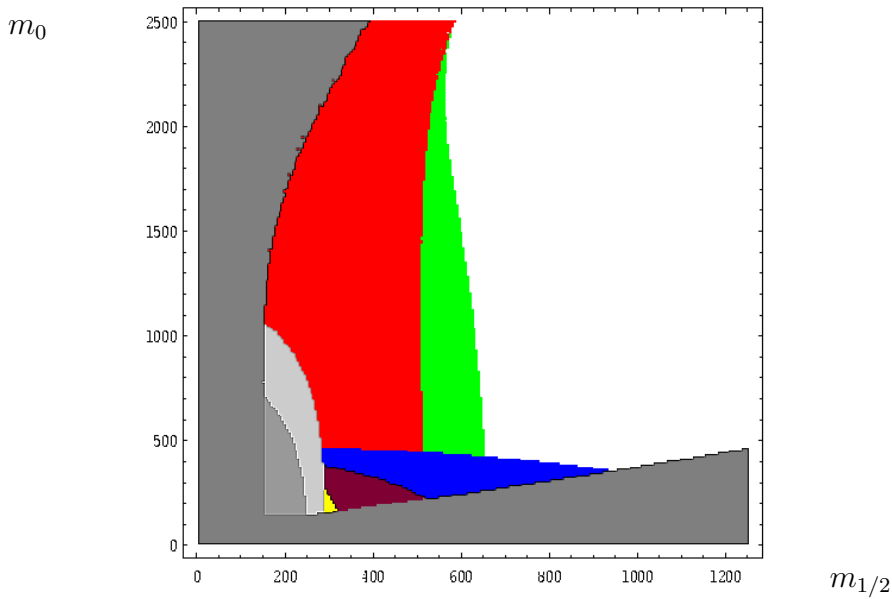,width=8.5cm}\\
\end{center}
\vspace*{-1cm}
\hspace*{12.3cm} {\large $m_{1/2}$}
\vspace*{1mm}
\caption[]{The $(m_{1/2}, m_0)$ mSUGRA plane with $\tan\beta=40,A_0 = 0$ and 
sign$(\mu)>0$ where SUSY and Higgs particles
can be produced at and $e^+e^-$ collider with a c.m. energy $\sqrt{s}=800$ 
GeV. The grey areas are those excluded by theoretical and experimental 
constraints. The colored regions are those where then cross sections are
large enough for the particles to be produced: $\tilde\chi_1^0 \tilde\chi_2^0$ 
(green), $\tilde\chi_1^+ \tilde\chi_1^-$ (red), $\tilde{l}^+ \tilde{l}^-$ 
(blue), $\tilde{\nu} \tilde{\nu}^*$ (purple), $\tilde{t}_1 \tilde{t}_1^*$ 
(dark blue) and the heavy MSSM $H,A,H^\pm$ bosons (yellow). Note that some of 
these regions are overlapping.} 
\vspace*{-.2cm}
\end{figure}

\section{Sparticle and Higgs production in $\ee$ Collisions}

\vspace*{-2mm}

We have then analyzed the prospects for producing SUSY particles and heavy
Higgs bosons at high--energy and high--luminosity $\ee$ colliders, requiring a
sample of 50 events per year to establish discovery; this should be sufficient
in the clean environment provided by $\ee$ colliders. At c.m. energies
$\sqrt{s} \sim 800$ GeV, typical of the TESLA machine [2], we have shown that
charginos, neutralinos and sleptons [in particular $\tilde{\tau}$ and
$\tilde{\nu}$] are accessible in rather large regions of the parameter space.
In particular, already at $\sqrt{s} = 800$ GeV associated $\lsp \tilde
\chi_2^0$ production is accessible in the entire overlap region described
above. Almost all of this region can also be probed through $\tilde \chi_1^\pm$
pair production, and in much of it $\tilde \tau_1$ pair production can also be
studied.  In some areas, top squarks and even bottom squarks can be produced.
In the large $\tb$ regime, where the present indications for SUSY can be
accommodated in a larger fraction of the $(m_{1/2}, m_0)$ plane, there is a
large region where the heavier MSSM Higgs bosons $H,A$ and $H^\pm$ are
kinematically accessible.  

Even for lower c.m. energies, $\sqrt{s} \sim 500$ GeV, charginos,
neutralinos and charged $(\tau)$ sleptons can be produced in a
significant region of parameter space not excluded by the present
constraints. However, discovery of sparticles can then no longer be
guaranteed [in the framework of mSUGRA] even if all positive
indications for SUSY hold up to further scrutiny. On the other hand,
if the c.m. energy of the collider is increased to $\sqrt{s}=1.2$ TeV,
the mSUGRA parameter space where SUSY and Higgs particles are
kinematically accessible and have sufficiently large cross sections to
be detected becomes very wide. The $\ee$ collider will then have a
search potential of SUSY particles that is comparable to the range
probed at the LHC. This is largely due to the fact that, thanks to the
high luminosities expected at future $\ee$ colliders, the process $\ee
\to \tilde\chi_1^0 \tilde\chi_2^0$ can probe large values of the
parameter $m_{1/2}$: only from kinematical arguments, values $m_{1/2}
\sim 1$ TeV can be probed at $\sqrt{s}=1.2$ TeV, corresponding to a
gluino mass of the order of 2 TeV. Heavy Higgs particles can be
searched if their masses are smaller than the beam energy. For large
values of $\tb$, this occurs in a large region of the mSUGRA parameter
space. 

Once these particles are found,  precision measurements at an $\ee$ collider
could reveal a great deal about the MSSM spectrum. In particular,
threshold scans allow the measurement of some sparticle masses at the permile
level. Making use of the ability to vary the beam polarization at will, various
couplings appearing in the production cross sections of SUSY and Higgs
particles can be measured with a high precision. Additional couplings can be
determined through the careful measurement of decay branching ratios. By
combining the information on sleptons and electroweak gauginos that one can
obtain at $\ee$ colliders with the information on squark and gluino production
obtained at the LHC would allow very stringent tests of the model.

\end{document}